\documentclass[12pt,aps,prd,preprint,tightenlines,superscriptaddress,
amsmath,amssymb,nofootinbib]{revtex4}
\RequirePackage[colorlinks=true
,urlcolor=blue
,anchorcolor=blue
,citecolor=blue
,filecolor=blue
,linkcolor=blue
,menucolor=blue
,pagecolor=blue
,linktocpage=true
,pdfproducer=medialab
,pdfa=true
]{hyperref}

\allowdisplaybreaks
\usepackage[utf8x]{inputenc}

\usepackage{amsmath,amssymb,amsthm,amsfonts}
\usepackage{graphicx}
\usepackage{subfigure}
\usepackage{dcolumn}	
\usepackage{hyperref}      	
\usepackage{bm}		
\usepackage{epsfig}
\usepackage{epstopdf}
\usepackage{setspace}
\usepackage[usenames, dvipsnames]{color}
\usepackage{slashed}
\usepackage{comment}
\usepackage{enumitem}

\newcommand{\PRE}[1]{{#1}} 
\usepackage{graphicx}
\usepackage{hyperref}
\usepackage[hyphenbreaks]{breakurl}
\usepackage{amsmath}
\usepackage{amssymb}
\usepackage{epsfig}
\usepackage{afterpage}
\usepackage{changepage}
\usepackage{amsmath}
\usepackage[normalem]{ulem}

\usepackage{placeins}
\usepackage{bm}

\def \SM{{\rm SM}}

\def \BtoXctaunu{\bar{B} \to X_c \tau^- \bar{\nu}_{\tau}}
\def \BtoXclnu{\bar{B} \to X_c \ell^- \bar{\nu}_{\ell}}

\def \RD{R({D^{(*)}})}

\def \({\left(}
\def \){\right)}
\def \[{\left[}
\def \]{\right]}
\def \l|{\left|}
\def \r|{\right|}

\def \qhs{\hat{q}^2}
\def \mth{\hat{m}_\tau}
\def \lam123{\lambda(1,\hat{q}^2,\rho^2)}

\def \pt{p_{\tau}}
\def \pn{p_{\nu}}

\def \SM{{\rm SM}}

\usepackage{bm}

\def \({\left(}
\def \){\right)}
\def \[{\left[}
\def \]{\right]}
\def \l|{\left|}
\def \r|{\right|}

\def \SM{{\rm SM}}

\def \RD{R({D^{(*)}})}

\def \({\left(}
\def \){\right)}
\def \[{\left[}
\def \]{\right]}
\def \l|{\left|}
\def \r|{\right|}

\begin{document}

\preprint{UMISS-HEP-2018-11}

\title{\PRE{\vspace*{1.0in}}
New physics in inclusive semileptonic $B$ decays including nonperturbative corrections
\PRE{\vspace*{.4in}}}

\author{Saeed Kamali}
\email{skamali@go.olemiss.edu}
\affiliation{Department of Physics and Astronomy, University of Mississippi, 108 Lewis Hall, Oxford, MS 38677 USA
\PRE{\vspace*{.5in}}}


\begin{abstract}
\PRE{\vspace*{.2in}}
In this work we study the effects of New Physics (NP) operators on the inclusive $\BtoXctaunu$ decay including power $(\mathcal{O}(1/m_b^2))$ corrections in the NP operators. In analogy with $R(D^{(*)})$ observables, we study the observable $R(X_c)=\frac{\mathcal{B}(\BtoXctaunu)}{\mathcal{B}(\BtoXclnu)}$. We present some numerical results for $R(X_c)$ and compare the results for this observable with and without power corrections in the NP contributions.    
\end{abstract}


\maketitle


\clearpage

\section{Introduction}
\label{sec:introduction}

Flavor anomalies have attracted a lot of attentions recently. Specially, the anomalies in the measurements of the $\bar{B} \to D^{(*)}$ transitions are interesting since they are confirmed by many experiments and have persisted for a long time. The measured quantities are the ratios of branching fractions of the semileptonic decays defined by $R(D^{(*)})=\mathcal{B}(\bar{B} \to D^{(*)} \tau^- \bar{\nu}_{\tau})/\mathcal{B}(\bar{B} \to D^{(*)} \ell^- \bar{\nu}_{\ell})$, where $\ell=e,\mu$ \cite{Lees:2013uzd, Lees:2012xj, Huschle:2015rga, Abdesselam:2016cgx, Sato:2016svk, Hirose:2016wfn, Aaij:2015yra}. These anomalies are rather robust since most of the experimental and theoretical uncertainties cancel in this ratio. They are interesting as these interactions happen at tree level and if approved, we need a large contribution from new physics (NP) to alleviate these deviations from theoretical predictions. There has been many studies of these anomalies in various NP models (see e.g.  \cite{Celis:2012dk, Duraisamy:2013kcw, Crivellin:2013wna, Dorsner:2013tla, Freytsis:2015qca, Deshpand:2016cpw, Bhattacharya:2018kig} and references there). Generically, these observables can be considered as tests of the lepton universality, so the assumed NP responsible for these anomalies should couple to leptons non-universally. Since the mass of the $\tau$ lepton is much larger than $\mu$ and $e$, and in view of the lepton flavor non-universality, we usually assume that NP only couples to the $\tau$ lepton \cite{Datta:2012qk, Duraisamy:2014sna, Bhattacharya:2014wla}, so it is present only in the $\BtoXctaunu$ decay. Here we follow the same approach and consider NP only in the third generation.\\

The $\SM$ predictions for $R(D)$ and $R(D^*)$ are (for $\ell=e$),
\begin{eqnarray}
\label{RD(s)_SM}
R(D)_{SM}=0.298 \pm 0.003, \nonumber\\
R(D^*)_{SM}=0.255 \pm 0.004.
\end{eqnarray}

There are lattice QCD predictions for the ratio $R(D)_{SM}$ in the Standard Model \cite{Bailey:2012jg, Lattice:2015rga, Na:2015kha} that are in good agreement with one another, 

\begin{eqnarray}
R(D)_{SM} &=& 0.299 \pm 0.011 \quad \quad [\mathrm{FNAL/MILC}], \\
R(D)_{SM} &=& 0.300\pm 0.008 \quad \quad [\mathrm{HPQCD}].
\end{eqnarray}

To calculate the $\SM$ predictions for $R(D)$ in Eq. (\ref{RD(s)_SM}), we use the results of \cite{Bigi:2016mdz} for the BGL parameterization where experimental and lattice results are used in the fit. There are also recent analyses of $\SM$ predictions of $R(D^*)$ \cite{Bigi:2017jbd, Bernlochner:2017jka, Jaiswal:2017rve}. To calculate the value for $R(D^*)_{SM}$ in Eq. (\ref{RD(s)_SM}), we use the results of the fit in \cite{Jaiswal:2017rve} for the CLN parameterization of the $B \to D^*$ form factors. These results are presented in table $4$ of this reference where they use the experimental data along with the lattice QCD and light cone sum rule results in the fit. \\
The averages of $R(D)$ and $R(D^*)$ measurements evaluated by the Heavy-Flavor Averaging Group are \cite{HFLAV16},

\begin{eqnarray}
R(D)_{exp}=0.407 \pm 0.039 \pm 0.024, \nonumber \\
R(D^*)_{exp}=0.306 \pm 0.013 \pm 0.007 .
\end{eqnarray}

These values exceed the $\SM$ predictions by more than $3\sigma$ \cite{HFLAV16}.\\

In view of these anomalies, it is logical to probe possible new physics effects in other decay modes which are connected to the $\RD$ anomalies via the same parton level transitions. An example of this kind of decay mode is the inclusive $\BtoXctaunu$ decay. In a recent work \cite{Kamali:2018fhr}, we studied effects of different NP Dirac structures on the inclusive decay $\BtoXclnu$. There, the NP contributions were considered at leading order. In this work we add the nonperturbative $1/m_b$ corrections to these NP Dirac structures and provide some numerical results for the effects of these corrections compared to the case when NP is added at parton level only.
In \cite{Grossman:1994ax} the inclusive $B$ decay is studied in the two higgs doublet model where a particular combination of the scalar and pseudoscalar couplings appear as NP contributions. In \cite{Colangelo:2016ymy}, nonperturbative corrections of order $\mathcal{O}(1/m_b^2)$ in the tensor currents are calculated. Here we present the results of these corrections for the scalar, pseudoscalar, vector and tensor contributions, including all the interference terms. This will help in a more precise study of the inclusive $\BtoXctaunu$ decay mode in the presence of NP.\\

In section \ref{sec:Inclusive B decay} we briefly describe the inclusive $B$ decay process and present the results of our calculations, in section \ref{sec:Results} we present the numerical results and in \ref{sec:Conclusions} we finish the note with a short conclusion.


\section{Inclusive B decay}
\label{sec:Inclusive B decay}
 
The inclusive semileptonic $B$ decay rate can be calculated systematically by expansion in terms of perturbative and nonperturbative corrections. The leading terms in this expansion reproduce the free quark decay rate while higher order terms are written as double expansions in terms of short distance perturbative effect which is an expansion in $\alpha_s$, and long distance nonperturbative  effect which is an expansion in $\Lambda_{QCD}/m_b$. 

 Nonperturbative corrections are calculated in the context of operator product expansion (OPE) and heavy quark effective theory (HQET). The techniques to calculate these corrections are known well (see e.g. \cite{Manohar:1993qn, Balk:1993sz, Falk:1993dh, Koyrakh:1993pq, Falk:1994gw, Blok:1993va, Ligeti:2014kia}).  The expansion is basically written in terms of operators with increasing dimensions where the higher dimension operators are suppressed by powers of $1/m_b$. A convenient method to calculate these corrections to arbitrary order in $1/m_b$, is presented in \cite{Dassinger:2006md}. In this note, we extend the $\SM$ results by adding the scalar, pseudo-scalar, vector and tensor currents as NP effects.
We consider the effective Hamiltonian, 

\begin{eqnarray}
\label{eq1:Ham}
 {\cal{H}}_{eff} &=&  \frac{G_F V_{cb}}{\sqrt{2}}\Big\{
\Big[\bar{c} \gamma_\mu (1-\gamma_5) b  + g_L \bar{c} \gamma_\mu (1-\gamma_5)  b + g_R \bar{c} \gamma_\mu (1+\gamma_5) b\Big] \bar{\tau} \gamma^\mu(1-\gamma_5) \nu_{\tau} \nonumber \\
				 && +  \Big[g_S\bar{c}  b   + g_P \bar{c} \gamma_5 b\Big] \bar{\tau} (1-\gamma_5)\nu_{\tau} + \Big[g_T\bar{c}\sigma^{\mu \nu}(1-\gamma_5)b\Big]\bar{\tau}\sigma_{\mu \nu}(1-\gamma_5)\nu_{\tau} + h.c. \Big\}, \nonumber \\ \label{eq:Heff}
\end{eqnarray}

where $G_F$ is the Fermi constant and $V_{cb}$ is the Cabibbo-Kobayashi-Maskawa (CKM) matrix element. When $g_S = g_P = g_L = g_R =g_T= 0$, the above equation produces the $\SM$ effective Hamiltonian.


To calculate the differential decay rate for $\BtoXctaunu$, we use the optical theorem to find the imaginary part of the time ordered products of the charged currents,

\begin{equation}
\int d^4x e^{-iq.x} \langle B |T\{\mathcal{O}^\dagger(x),\mathcal{O}(0)\}|B\rangle,
\end{equation}   

where $\mathcal{O}$ consists of SM and NP currents,
\begin{equation}
\mathcal{O}=(1+g_L)\bar{c}\gamma^\mu(1-\gamma_5)b + g_R \bar{c}\gamma^\mu(1+\gamma_5)b + g_S \bar{c}b + g_P \bar{c}\gamma_5 b + g_T \bar{c}\sigma^{\mu\nu}(1-\gamma_5)b.
\end{equation}

The time ordered product can then be written as an operator product expansion where a series of operators with increasing dimensions appear. Then, using the heavy quark effective theory, we can separate the residual momentum of the heavy quark in the hadron (which is of order $\Lambda_{QCD}$) and define the matrix elements of the nonrenomalizable operators in the operator expansion. This procedure leads to the determination of hadronic form factors. After contracting with the leptonic currents, we can calculate the three-fold differential decay rate $\frac{d\Gamma}{dq^2 dE_\tau dE_\nu}$. Here the kinematic variable $q^2$ is the dilepton invariant mass and $E_\tau$ and $E_\nu$ are the energies of the $\tau$ lepton and the corresponding neutrino in the rest frame of the $B$ meson. The explicit expression of the three-fold decay distribution in terms of the invariant quantities, is provided in the appendix.  
  
The leading order result is the free quark decay distribution and the first nonperturbative correction appears at order $\Lambda_{QCD}^2/m_b^2$. This correction is proportional to two hadronic parameters $\lambda_1$ and $\lambda_2$ (or $\mu_\pi^2$ and $\mu_G^2$) which correspond to the kinetic energy and the spin interaction energy of the $b$ quark in the hadron, respectively.\\

After integrating over the energies of the charged lepton and the neutrino, we can find the $q^2$ distribution as,

\begin{align}
\frac{d\Gamma}{d\qhs}=& N(\qhs) \Big[ (|1+g_L|^2+|g_R|^2)\frac{d\Gamma}{d\qhs}\bigg|_{SM} + Re(g_R^*(1+g_L))\frac{d\Gamma}{d\qhs}\bigg|_{LR} + |g_S|^2\frac{d\Gamma}{d\qhs}\bigg|_{S} \nonumber\\[8pt]
                    &   + Re(g_S^*(1+g_L+g_R))\frac{d\Gamma}{d\qhs}\bigg|_{SLR} + |g_P|^2\frac{d\Gamma}{d\qhs}\bigg|_{P} + Re(g_P^*(1+g_L-g_R))\frac{d\Gamma}{d\qhs}\bigg|_{PLR} \nonumber\\[8pt]
                    &  +|g_T|^2 \frac{d\Gamma}{d\qhs}\bigg|_{T} + Re((1+g_L)g_T^*) \frac{d\Gamma}{d\qhs}\bigg|_{LT} + Re(g_R g_T^*) \frac{d\Gamma}{d\qhs}\bigg|_{RT}  \Big],
\end{align}

where $N(\qhs)=\frac{G_F^2 |V_{cb}|^2m_b^5(1-\mth^2/\qhs)^2}{96 \pi^3 \sqrt{\lam123}}$ and $\lambda(a,b,c)=a^2+b^2+c^2-2ab-2ac-2bc$. The various terms on the right hand side of the above equation are presented in the following, with subscripts that correspond to contributions of SM, NP and interference terms,

\begin{align}
\frac{d\Gamma}{d\qhs}\bigg|_{SM}=&  \Big(1+\frac{\lambda_1}{2m_b^2} \Big) \lam123 \Big\{ \big[ (1-\rho)^2+\qhs(1+\rho)-2(\qhs)^2 \big]  \nonumber \\
 								 & +\frac{\mth^2}{\qhs} \big[ 2(1-\rho)^2-\qhs(1+\rho)-(\qhs)^2 \big]  \Big\}  +\frac{3\lambda_2}{2m_b^2} \Big\{ \big[ (1-\rho)^3(1-5\rho)-\qhs(1-\rho)^2(1+5\rho) \nonumber \\
								& -3(\qhs)^2(5+6\rho+5\rho^2)+25(\qhs)^3(1+\rho)-10(\qhs)^4 \big] \nonumber\\ 
 & +\frac{\mth^2}{\qhs} \big[ 2(1-\rho)^3(1-5\rho)-\qhs(5-9\rho-21\rho^2+25\rho^3) \nonumber\\
 & +3(\qhs)^2(1+2\rho+5\rho^2)+5(\qhs)^3(1+\rho)-5(\qhs)^4 \big] \Big\}, 
\end{align}

\begin{align}
\frac{d\Gamma}{dq^2}\bigg|_{LR}=&~     -12 \sqrt{\rho}\qhs \Big(1+\frac{\lambda_1}{2m_b^2} \Big)\lam123 + 4\sqrt{\rho} \frac{3\lambda_2}{2m_b^2} \Big\{ \big[  2(1-\rho)^3-3\qhs(1-\rho)^2 \nonumber\\
								&+12(\qhs)^2(1+\rho)-7(\qhs)^3  \big] + \frac{4\mth^2}{\qhs} \big[ (1-\rho)^3-3\qhs\rho(1-\rho)-3\rho(\qhs)^2+(\qhs)^3 \big]     \Big\},
\end{align}

\begin{align}
\frac{d\Gamma}{d\qhs}\bigg|_{S}=& ~\frac{3\qhs}{4}((1+\sqrt{\rho})^2-\qhs) \Big[ \Big(1+\frac{\lambda_1}{2m_b^2}\Big)\lam123 \nonumber \\
 & +\frac{3\lambda_2}{2m_b^2} \Big( (1-\sqrt{\rho})^2(1+6\sqrt{\rho}+5\rho)-2\qhs(1-2\sqrt{\rho}+5\rho)+5(\qhs)^2  \Big)  \Big], 
\end{align}

\begin{align}
\frac{d\Gamma}{d\qhs}\bigg|_{SLR}=& ~\frac{3\mth}{2}(1-\sqrt{\rho})((1+\sqrt{\rho})^2-\qhs) \Big[  \Big(1+\frac{\lambda_1}{2m_b^2} \Big)\lam123 \nonumber\\
 &  + \frac{3\lambda_2}{2m_b^2} \Big( (1-\sqrt{\rho})^2(1+6\sqrt{\rho}+5\rho)-2\qhs(1-2\sqrt{\rho}+5\rho)+5(\qhs)^2  \Big)  \Big], 
\end{align}

\begin{align}
\frac{d\Gamma}{d\qhs}\bigg|_{P}=& ~ \frac{3\qhs}{4}((1-\sqrt{\rho})^2-\qhs) \Big[  \Big(1+\frac{\lambda_1}{2m_b^2}\Big)\lam123  \nonumber \\
								& +\frac{3\lambda_2}{2m_b^2} \Big( (1+\sqrt{\rho})^2(1-6\sqrt{\rho}+5\rho)-2\qhs(1+2\sqrt{\rho}+5\rho)+5(\qhs)^2   \Big)   \Big],  
\end{align}

\begin{align}
\frac{d\Gamma}{d\qhs}\bigg|_{PLR}=& ~\frac{3\mth}{2}(1+\sqrt{\rho})((1-\sqrt{\rho})^2-\qhs) \Big[ \Big(1+\frac{\lambda_1}{2m_b^2}\Big)\lam123  \nonumber\\
 & +\frac{3\lambda_2}{2m_b^2} \Big( (1+\sqrt{\rho})^2(1-6\sqrt{\rho}+5\rho)-2\qhs(1+2\sqrt{\rho}+5\rho)+5(\qhs)^2  \Big)   \Big], 
\end{align}

\begin{align}
\frac{d\Gamma}{d\qhs}\bigg|_{T}=& ~8(1+\frac{2\mth^2}{\qhs}) \Big[ \Big(1+\frac{\lambda_1}{2m_b^2} \Big) \Big(  2(1-\rho)^4-5\qhs(1-\rho)^2(1+\rho) +(\qhs)^2(3+2\rho+3\rho^2)\nonumber\\
								& +(\qhs)^3(1+\rho)-(\qhs)^4 \Big) + \frac{3\lambda_2}{2m_b^2} \Big( 2(-1+\rho)^3(3+5\rho)+\qhs(3+17\rho+5\rho^2-25\rho^3)\nonumber\\
								 &+(\qhs)^2(3+14\rho+15\rho^2)+5(\qhs)^3(1+\rho)-5(\qhs)^4    \Big)       \Big],
\end{align}

\begin{align}
\frac{d\Gamma}{d\qhs}\bigg|_{LT}=& ~36\mth\sqrt{\rho} \Big[\Big(1+\frac{\lambda_1}{2m_b^2} \Big)\Big( (-1+\rho)^3+\qhs(1+2\rho-3\rho^2)+(\qhs)^2(1+3\rho)-(\qhs)^3  \Big) \nonumber\\
							&+\frac{\lambda_2}{2m_b^2} \Big( (1-\rho)^2(1+15\rho)+\qhs(3+10\rho-45\rho^2)+(\qhs)^2(19+45\rho)-15(\qhs)^3  \Big)      \Big],
\end{align}

\begin{align}
\frac{d\Gamma}{d\qhs}\bigg|_{RT}=& ~-36\mth \Big[\Big(1+\frac{\lambda_1}{2m_b^2} \Big)\Big( (-1+\rho)^3-\qhs(-3+2\rho+\rho^2)-(\qhs)^2(3+\rho)+(\qhs)^3     \Big) \nonumber\\
								& + \frac{\lambda_2}{2m_b^2} \Big( (1-\rho)^2(5+11\rho)+\qhs(1-18\rho-15\rho^2)-(\qhs)^2(13+3\rho)+7(\qhs)^3    \Big)       \Big].
\end{align}

Here we have defined the normalized quantities, $\qhs=q^2/m_b^2$, $\rho=m_c^2/m_b^2$ and $\mth=m_\tau/m_b$. Note that there is no scalar-pseudoscalar or (pseudo)scalar-tensor interference terms in the $q^2$ distribution.
For $g_S=g_P=g_L=g_R=g_T=0$, we reproduce the SM results and for $g_S=g_P=g_L=g_R=0$ we reproduce the results given in \cite{Colangelo:2016ymy}.

\section{Numerical Results}
\label{sec:Results}
In this section we present the numerical results of our calculations in two mass schemes for the quarks masses: the $1S$ mass scheme \cite{Hoang:1998ng, Hoang:1998hm} and the kinetic scheme \cite{Benson:2003kp, Gambino:2004qm, Gambino:2011cq, Alberti:2014yda}. In the $1S$ scheme, we follow  \cite{Bauer:2002sh, Bauer:2004ve} to write the rate in terms of the nonperturbative parameters, $m_b$, $\lambda_1$ at $\mathcal{O}(1/m_b^2)$ and $\rho_1$, $\tau_1$ and $\tau_3$ at $\mathcal{O}(1/m_b^3)$, and we use the numerical results of the fit together with the correlations between the parameters from \cite{HFLAV16}. In the kinetic scheme the nonperturbative parameters are $m_b$ and $m_c$, $\mu^2_{\pi}$ and  $\mu_G^2$ at $\mathcal{O}(1/m_b^2)$ and $\rho_D^3$ at $\mathcal{O}(1/m_b^3)$. The numerical values of these parameters together with their correlation matrix are presented in \cite{Alberti:2014yda,Bhattacharya:2018kig}. We present the numerical inputs in table \ref{Table:numbers}. The correlation matrices of these parameters are taken from the references mentioned in the table and we do not repeat them here.

\begin{table}[]
\label{Table:numbers}
\begin{tabular}{|c|c||c|c|}
\hline
Parameter        & Value \cite{HFLAV16}          &     Parameter              & Value \cite{Bhattacharya:2018kig} \\
($1S$ scheme)    &                               &    (kinetic scheme)        &                                                    \\ \hline
$m_b^{1S}$       & $4.691 \pm 0.037~GeV$         & $m_{b}^{kin}$              & $4.561 \pm 0.021~GeV$          \\ \hline
$\lambda_1$      & $-0.362 \pm 0.067~GeV^2$      & $m_c$                      & $1.092 \pm 0.020~GeV$          \\ \hline
$\rho_1$         & $0.043 \pm 0.048~GeV^3$       & $\mu_\pi^2$                & $0.464 \pm 0.067~GeV^2$        \\ \hline
$\tau_1$         & $0.161 \pm 0.122~GeV^3$       & $\rho_D^3$                 & $0.175 \pm 0.040~GeV^3$        \\ \hline
$\tau_3$         & $0.213 \pm 0.102~GeV^3$       & $\mu_G^2$                  & $0.333 \pm 0.061~GeV^2$        \\ \hline
\end{tabular}
\caption{Values of the parameters used for the numerical results. The correlation matrices are taken from the references mentioned in the table.}
\end{table}

In our numerical results we also include the $\mathcal{O}(1/m_b^3)$ correction in SM which is derived in \cite{Mannel:2017jfk}. Besides nonperturbative effects, we include the $\mathcal{O}(\alpha_s)$ perturbative corrections in $\SM$ calculated in \cite{Aquila:2005hq, Jezabek:1996db}. The effects of higher order perturbative corrections are very small in the observables where the ratio of rates are calculated \cite{Biswas:2009rb, Kamali:2018fhr}, so we include only $\mathcal{O}(\alpha_s)$ corrections.

We find for the ratio of branching ratios in $\SM$, $R(X_c)_{SM}=\frac{\mathcal{B}(B \to X_c \tau^- \bar{\nu}_{\tau})_{SM}}{\mathcal{B}(B \to X_c \ell^- \bar{\nu}_{\ell})_{SM}},$ in the $1S$ scheme,

\begin{equation}
R(X_c)_{SM}^{1S}=0.216 \pm 0.003 ~,
\end{equation}
 and in the kinetic scheme,
 
\begin{equation}
R(X_c)_{SM}^{kin}=0.213 \pm 0.004~.
\end{equation}

 Adding the NP effects, we can find in the $1S$ scheme,
\begin{align}
\frac{R(X_c)_{~~}^{1S}}{R(X_c)_{SM}^{1S}}\simeq &~ 1+ 1.147 \big( |g_L|^2+|g_R|^2+2Re(g_L) \big) + 0.031|g_P|^2 + 0.327|g_S|^2 + 12.637|g_T|^2 \nonumber\\
								 &  - 0.714Re((1+g_L)g_R^*) + 0.096 Re((1+g_L-g_R)g_P^*) + 0.493Re((1+g_L+g_R)g_S^*)\nonumber \\
								 &  + 5.514Re(g_Rg_T^*) - 3.402 Re((1+g_L)g_T^*),
\end{align}  

and similarly in the kinetic scheme,

\begin{align}
\frac{R(X_c)_{~~}^{kin}}{R(X_c)_{SM}^{kin}}\simeq &~ 1+ 1.266 \big( |g_L|^2+|g_R|^2+2Re(g_L) \big) + 0.042|g_P|^2 + 0.351|g_S|^2 + 13.969|g_T|^2 \nonumber\\
								 &  - 0.744Re((1+g_L)g_R^*) + 0.120 Re((1+g_L-g_R)g_P^*) + 0.525Re((1+g_L+g_R)g_S^*)\nonumber \\
								 &  + 6.094Re(g_Rg_T^*) - 3.462 Re((1+g_L)g_T^*).
\end{align}

There is a measurement of the inclusive rate by ALEPH  \cite{Barate:2000rc},

\begin{equation}
\mathcal{B}(b \to X \tau^-\bar{\nu}_\tau)_{exp}=(2.43 \pm 0.32)\times 10^{-2}
\end{equation}

where $X=X_c + X_u$ are all possible states from $b \to c$ and $b \to u$ transitions. This measurement is dominated by the $b \to c$ mode since $\frac{|V_{ub}|}{|V_{cb}|}=0.083\pm0.006$ as measured by LHCb \cite{Aaij:2015bfa}. On the other hand the $b \to u$ mode has a larger phase space compared to the $b \to c$ mode. We estimate the contribution of the $b \to u$ mode to this measurement by,

\begin{equation}
\mathcal{B}(b \to X \tau^-\bar{\nu}_\tau)_{exp} \approx \mathcal{B}(b \to X_c \tau^-\bar{\nu}_\tau)_{exp}(1 + \frac{|V_{ub}|^2}{|V_{cb}|^2}\times 2.8),
\end{equation} 
where the factor $2.8$ is due to the larger phase space in the $b \to u$ mode. This estimation which is consistent with the one given in \cite{Celis:2016azn} leads to,
\begin{equation}
 \mathcal{B}(b \to X_c \tau^-\bar{\nu}_\tau)_{exp}=(2.38 \pm 0.32)\times 10^{-2}.
\end{equation}

Note that the ALEPH measurement represents the inclusive weak decay for a mixture of $b$ hadrons and to leading order in the heavy quark expansion, all $b$ hadrons have the same width. So this measurement can be considered as the branching ratio for each individual $b$ hadron. 

Using the world average for the semileptonic branching ratio into the light lepton \cite{HFLAV16}, $\mathcal{B}(B \to X_c \ell^- \bar{\nu}_{\ell})_{exp}=(10.65 \pm 0.16)\times 10^{-2}$, we can find an experimental value for the ratio,
\begin{equation}
R(X_c)_{exp}=0.223 \pm 0.030.
\end{equation}

\begin{figure}
\begin{center}
\includegraphics[width=5.5cm]{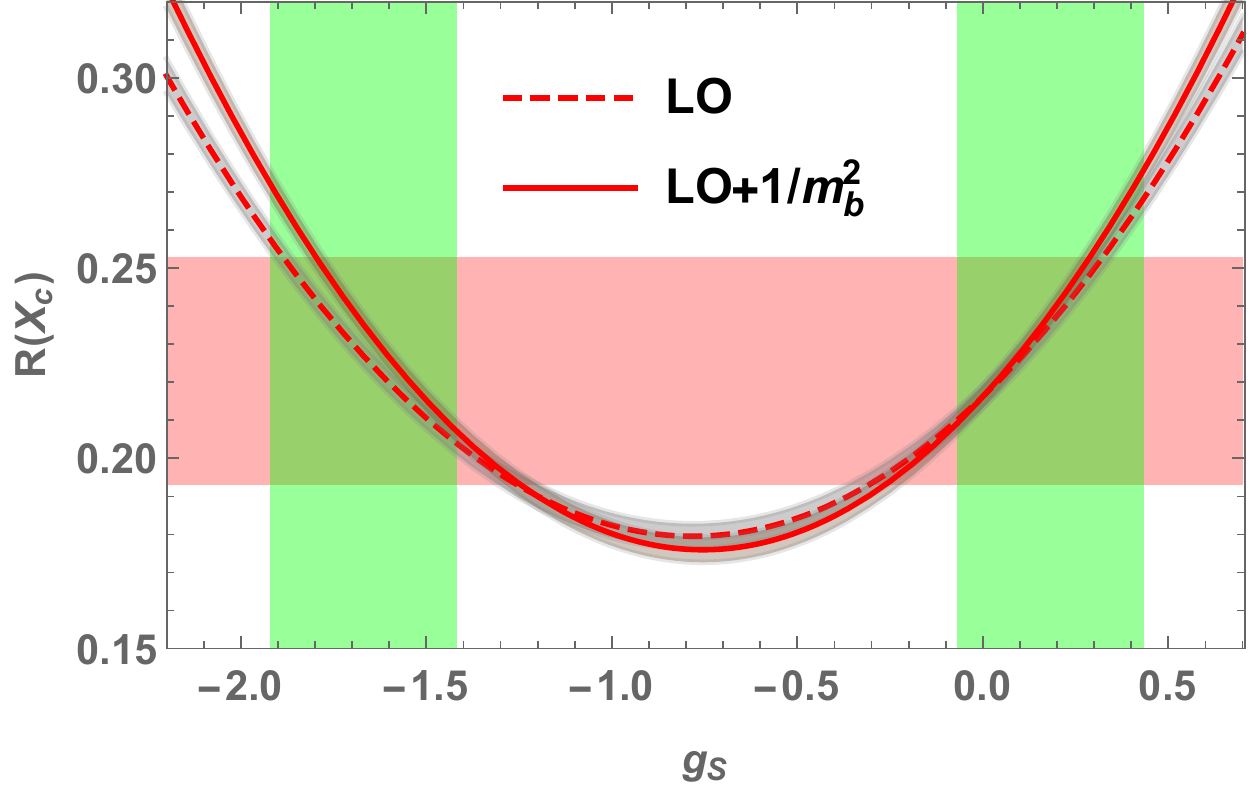}~~~
\includegraphics[width=5.5cm]{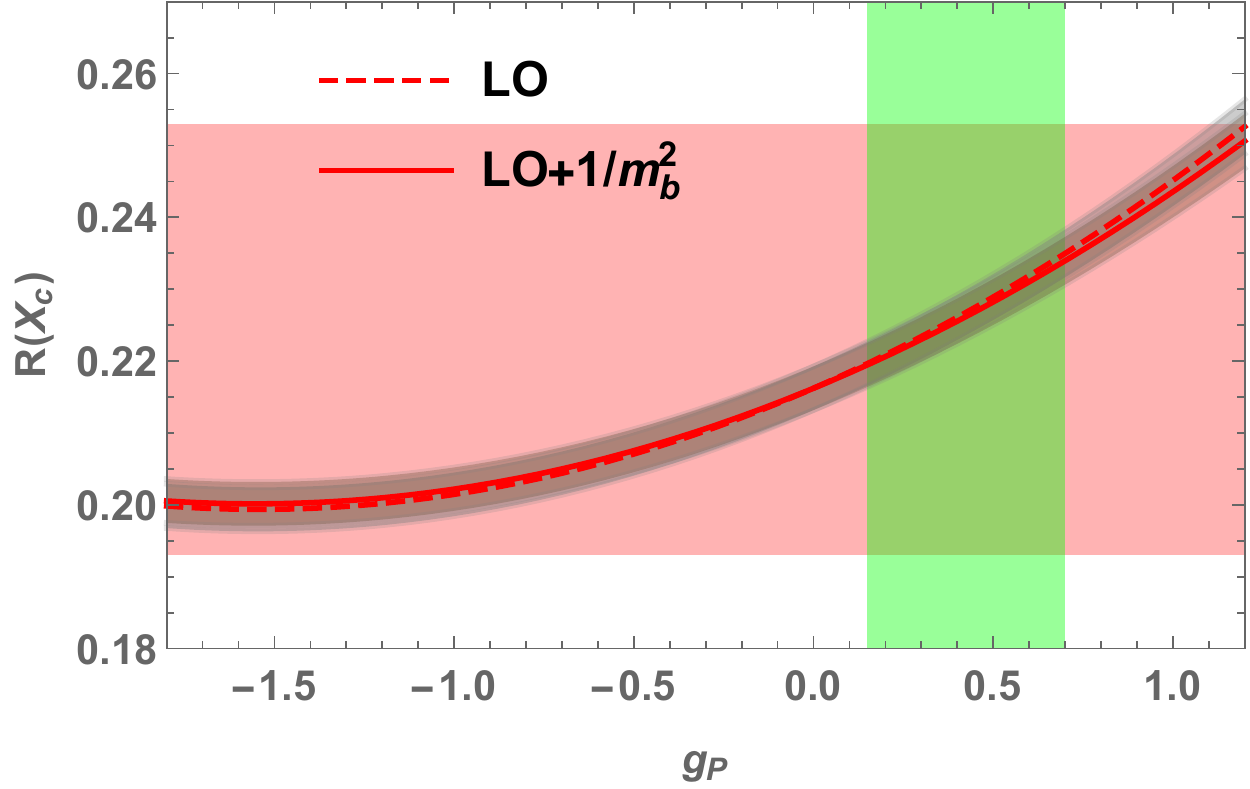}~~~\\[0.3cm]
\end{center}
\begin{adjustwidth}{-0.5cm}{-0.5cm}
\includegraphics[width=5.5cm]{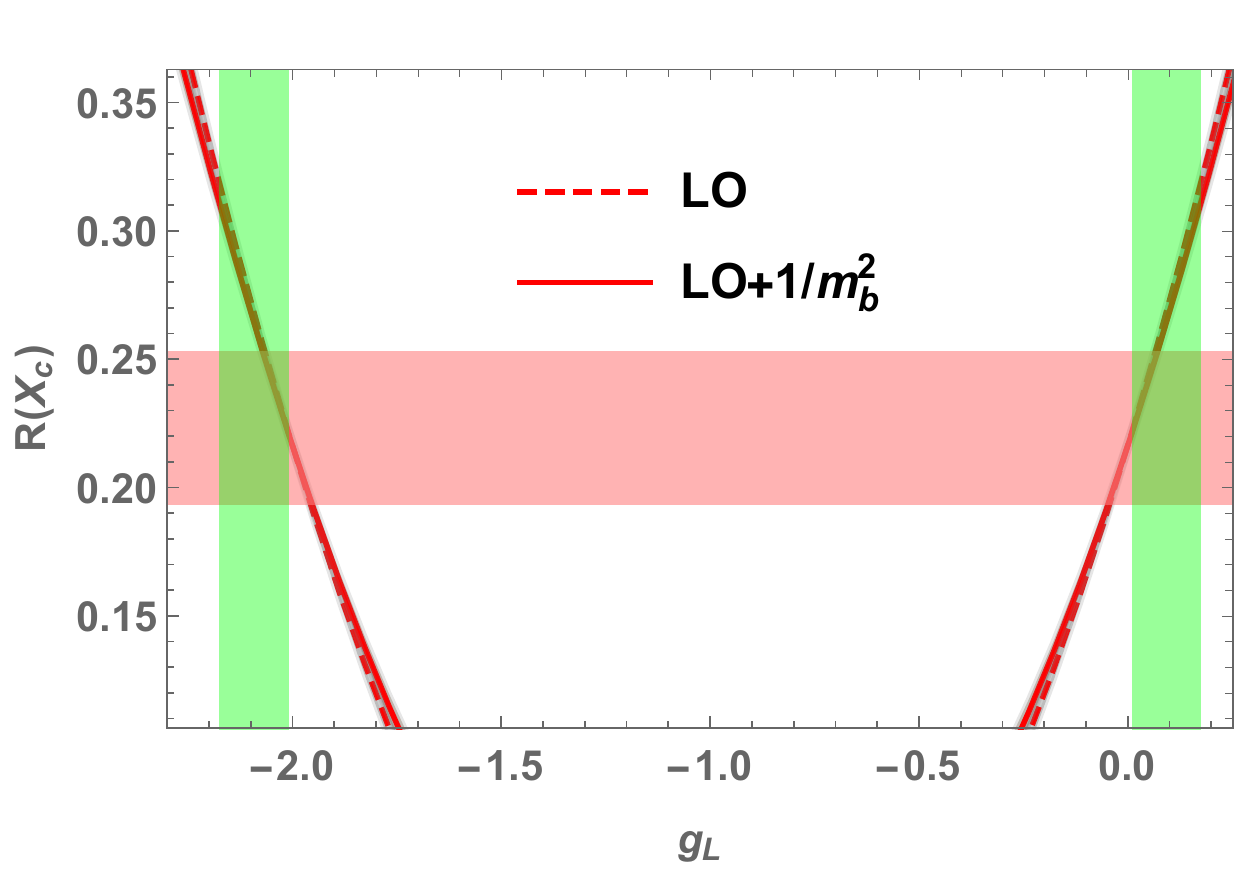}~~
\includegraphics[width=5.5cm]{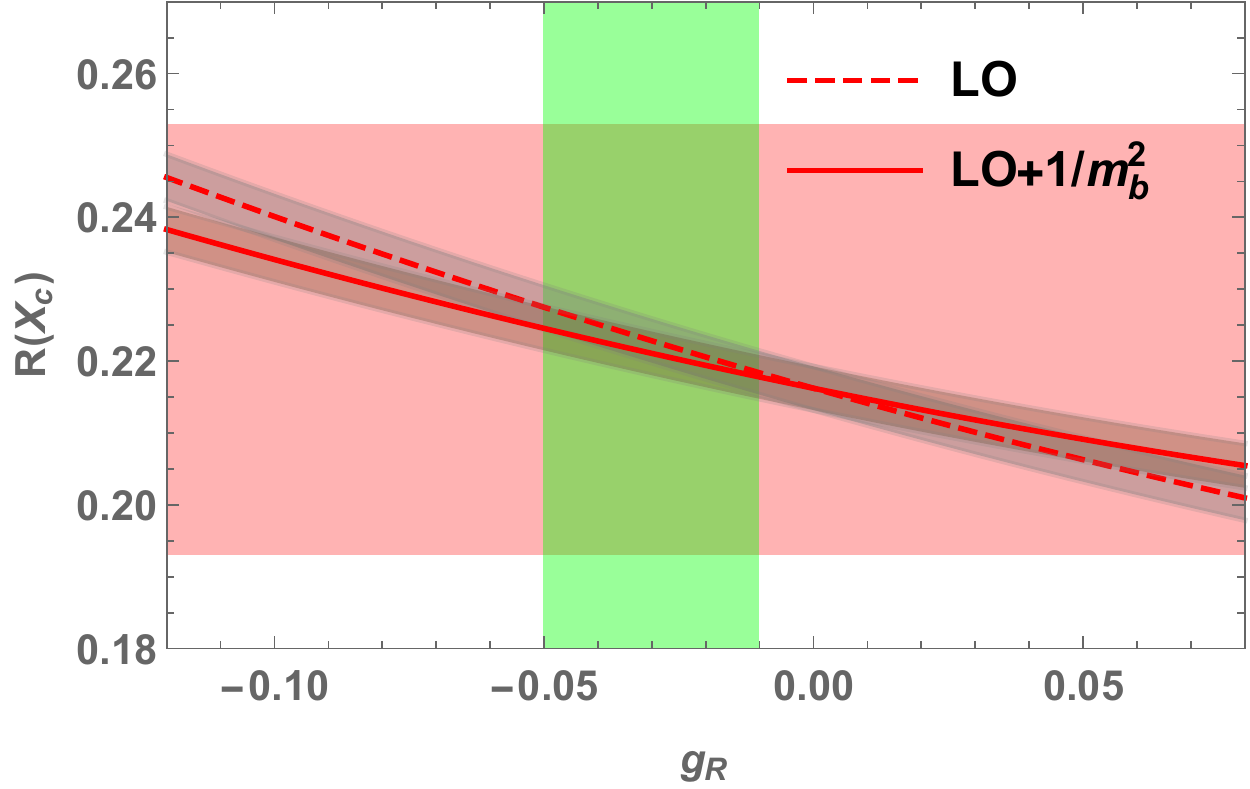}~~
\includegraphics[width=5.5cm]{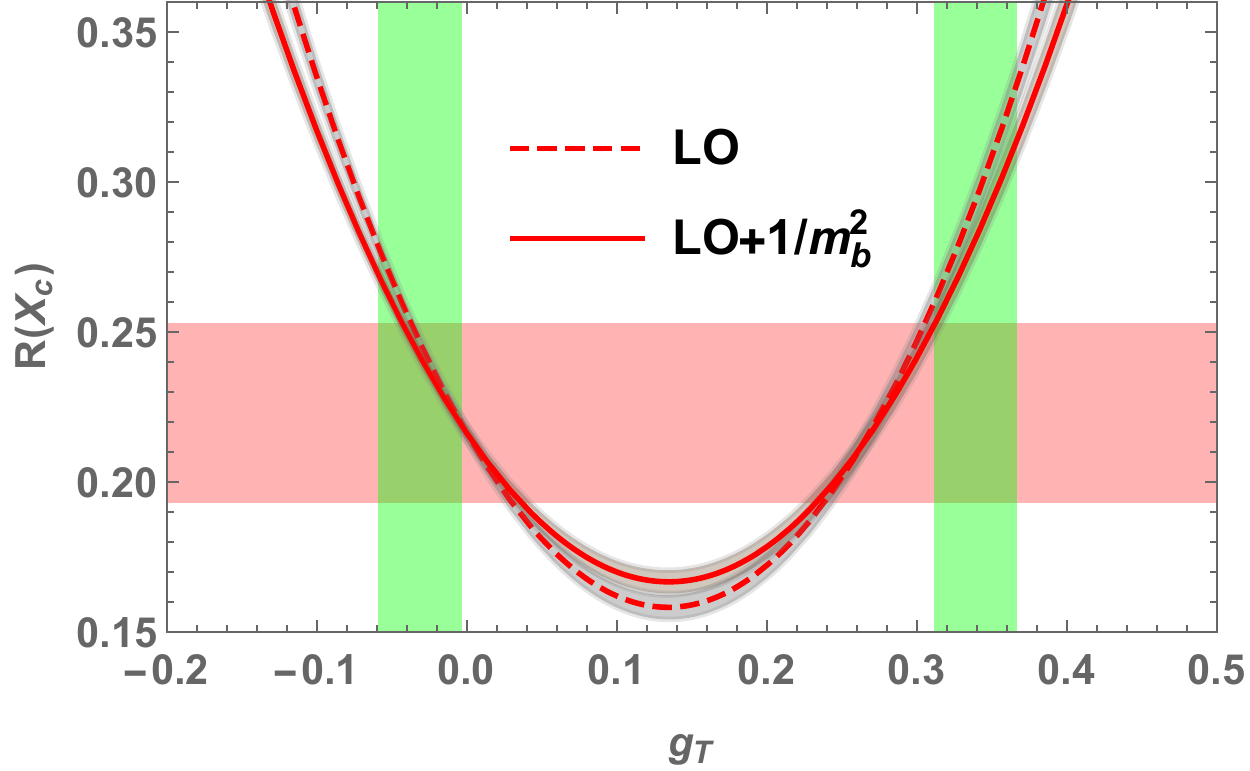}
\end{adjustwidth}
\caption{The ratio of decay rates $R(X_c)$ (in $1S$ scheme) when one coupling at a time is present. The dashed red curves correspond to the case when the NP contribution is added at parton level while the solid red curves correspond to the case when power corrections are included in the NP contributions. Green bands are the constraints on the couplings due to $R(D^{(*)})_{exp}$ within $3\sigma$ and $B_c$ lifetime. The pink band is $R(X_c)_{exp}$ within $1\sigma$.}
\label{fig_RXc}
\end{figure}

In Fig. (\ref{fig_RXc}) we present the results (in the $1S$ scheme) for the observable $R(X_c)$ when we turn on one NP coupling at a time. We consider two cases: the first case is when the NP contribution is considered only at parton level(dashed red curves), and the second case is when we add the subleading $1/m_b$ corrections to these NP contributions(solid red curves). The gray and brown bands correspond to the uncertainties of this observable when we vary the values of the parameters within their uncertainties. The green bands are the constraints on the couplings when we consider the measurements of $R(D^{(*)})$ within $3\sigma$. For the $g_P$ coupling, it is well known that the $B_c$ lifetime leads to a tight constraint \cite{Li:2016vvp, Alonso:2016oyd, Celis:2016azn}. We use $\mathcal{B}(B_c \to \tau^- \bar{\nu}_\tau)\leq 30\%$ as in \cite{Datta:2017aue}, to include this constraint on the $g_P$ coupling which is included in the green band in the plot. The pink band, is the value of $R(X_c)_{exp}$ within $1\sigma$.\\
In the parameter space of interest, adding the $1/m_b$ corrections to the NP contributions causes a change of $R(X_c)$ that is numerically at percent level. This change is mostly noticeable in the $g_S$ and $g_T$ case where the maximum correction, in the parameter space that is favored by $R(D^{(*)})$, is $\approx 5\%$.

\section{Conclusions}
\label{sec:Conclusions}
Recent measurements of $R(D^{(*)})$ show large deviations from $\SM$ predictions and this could be a signal of nonuniversal NP. The quark level transition in this observable is $b \to c \tau^- \bar{\nu}_\tau$ and we can probe this transition in other decay modes. 
In a recent work \cite{Kamali:2018fhr}, we studied the inclusive $\BtoXctaunu$ decay in view of the anomalies in the $R(D^{(*)})$ measurements. In this work we extended this study by including the effects of $1/m_b$ corrections in the NP Dirac structures. We presented the results of our calculations for the differential decay rate $\frac{d\Gamma}{dq^2}$ as well as the three-fold decay distribution and presented some numerical results of the effects of these power corrections on the observable $R(X_c)$. By constraining the NP parameters by the existing $R(D^{(*)})$ measurements, we presented the favored parameter region by these measurements to illustrate if the power corrections in the NP part are important. We found that, in the parameter range of interest, these corrections are generically at percent level (except for the $g_P$ coupling which is smaller) and the maximum effect of these corrections is in the $g_S$ and $g_T$ part which is $\approx 5\%$.

\newpage
\acknowledgments
This work was supported in part by the National Science Foundation under Grant No.\ 
PHY-1414345, and in part by the dissertation fellowship from the graduate school and the summer research assistantship from the College of Liberal Arts at the University of Mississippi. The author acknowledges fruitful comments and discussions with A. Datta and Z. Ligeti. The author also acknowledges the hospitality of the Department of Physics and Astronomy at the University of California, Irvine, where part of the work was done. 

\appendix
\section{The three-fold differential distribution}
\label{appendix: diff_rate}
In this appendix we present the three-fold differential rate in terms of invariant quantities. We write the distribution in the presence of all NP couplings in the form,

\begin{align}
\frac{d^3\Gamma}{dx^3}=&~ \frac{G_F^2 |V_{cb}|^2}{8\pi^3} \Big\{ |1+g_L|^2 \frac{d^3\Gamma}{dx^3}\bigg|_{SM}  + |g_R|^2 \frac{d^3\Gamma}{dx^3}\bigg|_{R}  + |g_S|^2 \frac{d^3\Gamma}{dx^3}\bigg|_{S} + |g_P|^2 \frac{d^3\Gamma}{dx^3}\bigg|_{P} +|g_T|^2 \frac{d^3\Gamma}{dx^3}\bigg|_{T} \nonumber\\[8pt]
						& + Re((1+g_L)g_R^*) \frac{d^3\Gamma}{dx^3}\bigg|_{LR} + Re((1+g_L+g_R)g_S^*) \frac{d^3\Gamma}{dx^3}\big|_{SLR} \nonumber \\[8pt]
						& + Re((1+g_L-g_R)g_P^*) \frac{d^3\Gamma}{dx^3}\bigg|_{PLR} + Re((1+g_L)g_T^*) \frac{d^3\Gamma}{dx^3}\bigg|_{LT} + Re(g_Rg_T^*) \frac{d^3\Gamma}{dx^3}\bigg|_{RT} \nonumber\\[8pt]
						& + Re((g_S - g_P)g_T^*) \frac{d^3\Gamma}{dx^3}\bigg|_{SPT}   \Big\},
\end{align}

where the three independent variables are usually taken to be $dx^3=dq^2dE_{\tau}dE_{\nu}$ or $dx^3=dq^2dE_{\tau}dq.v$, $v$ being the four velocity of the $B$ meson. Each contribution to the differential rate can be written as,

\begin{equation}
\label{eq: decay_distrib}
\frac{d^3\Gamma}{dx^3}\bigg|_{A}=\frac{1}{\Delta_0}\frac{d^3\Gamma}{dx^3}\bigg|_{A}^{(1)} + \frac{1}{\Delta_0^2}\frac{d^3\Gamma}{dx^3}\bigg|_{A}^{(2)}+\frac{1}{\Delta_0^3}\frac{d^3\Gamma}{dx^3}\bigg|_{A}^{(3)}.
\end{equation} 

Here we have defined $\Delta_0=p^2-m_c^2$ with $p=m_b v - q$. The contributions (\ref{eq: decay_distrib}) to the decay distribution are given by the substitutions \cite{Manohar:1993qn, Balk:1993sz},

\begin{align}
\frac{1}{\Delta_0} 		\quad  &\to  \quad   \delta(p^2 - m_c^2) \nonumber\\
\frac{1}{\Delta_0^2}	\quad  &\to  \quad   -\delta^{\prime}(p^2 - m_c^2) \nonumber\\
\frac{1}{\Delta_0^3}	\quad  &\to	 \quad   \frac{1}{2}\delta^{\prime\prime}(p^2 - m_c^2). 
\end{align}  

In the following we present various contributions to this distribution.\\
The $\SM$ contribution is given as,

\begin{align}
\frac{d^3\Gamma}{dx^3}\bigg|_{SM}^{(1)}=&~ \frac{4}{3m_b}\big[ 6m_b p.\pt \pn.v+(\lambda_1+3\lambda_2)(2\pt .\pn - 5 \pt.v \pn.v)     \big] \\[8pt]
\frac{d^3\Gamma}{dx^3}\bigg|_{SM}^{(2)}=&~ \frac{4}{3m_b}\big[ 2(\lambda_1+3\lambda_2)(-2p.\pn+5p.v\pn.v)p.\pt + 2m_b\lambda_1(2p.v\pt.v-5p.\pt)\pn.v \nonumber \\
										& + 6m_b \lambda_2(p.v\pt.\pn-p.\pn\pt.v)      \big] \\[8pt]
\frac{d^3\Gamma}{dx^3}\bigg|_{SM}^{(3)}=&~ \frac{32\lambda_1}{3}    \big[ p.p-(p.v)^2 \big]  p.\pt \pn.v  ~     .									
\end{align}

The $A=R$ contribution is derived from $\SM$ part by the substitutions $\pt \to \pn$ and $\pn \to \pt$,  
\begin{align}
\frac{d^3\Gamma}{dx^3}\bigg|_{R}^{(i)} = ~\frac{d^3\Gamma}{dx^3}\bigg|_{SM}^{(i)}(\pt \leftrightarrow \pn) \quad \quad i=1,2,3 .
\end{align} 

For $A=S$ we have,

\begin{align}
\frac{d^3\Gamma}{dx^3}\bigg|_{S}^{(1)} =& ~\frac{1}{2m_b^2}\big[ 2m_b^2(p.v+m_c) + (m_b+m_c)(\lambda_1+3\lambda_2)     \big] \pt.\pn \\[8pt]
\frac{d^3\Gamma}{dx^3}\bigg|_{S}^{(2)} =& ~-\frac{(\lambda_1+3\lambda_2)}{3m_b} \big[ 3m_b(p.v+m_c)-3m_c p.v + 2p.p - 5(p.v)^2   \big]\pt.\pn  \\[8pt]
\frac{d^3\Gamma}{dx^3}\bigg|_{S}^{(3)} =& ~\frac{4\lambda_1 }{3}(p.v+m_c)\big[ p.p -(p.v)^2     \big] \pt.\pn ~,
\end{align}

while the $A=P$ case can be derived from $A=S$ case by the substitution $m_c \to -m_c$,
\begin{align}
\frac{d^3\Gamma}{dx^3}\bigg|_{P}^{(i)} =& ~\frac{d^3\Gamma}{dx^3}\bigg|_{S}^{(i)}(m_c \to - m_c) \quad \quad i=1,2,3.
\end{align}

For $A=T$ we find,
\begin{align}
\frac{d^3\Gamma}{dx^3}\bigg|_{T}^{(1)}=& ~\frac{16}{3m_b}\big[ 6m_b(2p.\pn \pt.v + 2p.\pt \pn.v - p.v \pt.\pn) + 5(\lambda_1+3\lambda_2)(\pt.\pn - 4\pt.v \pn.v)       \big] \\[8pt]
\frac{d^3\Gamma}{dx^3}\bigg|_{T}^{(2)}=& ~-\frac{32}{3m_b}\big[ (\lambda_1+3\lambda_2)(8p.\pt p.\pn-2p.p\pt.\pn+5(p.v)^2\pt.\pn-10p.\pn p.v \pt.v \nonumber\\
										& -10p.\pt p.v \pn.v  ) + 2m_b (5\lambda_1-3\lambda_2)(p.\pn \pt.v+p.\pt \pn.v)  \nonumber \\
										& -3m_b(\lambda_1-\lambda_2)p.v\pt.\pn - 8m_b\lambda_1 p.v\pt.v\pn.v    \big] \\[8pt]
\frac{d^3\Gamma}{dx^3}\bigg|_{T}^{(3)}=& ~-\frac{128\lambda_1}{3}\big[ p.p - (p.v)^2  \big] \big[ p.v \pt.\pn -2p.\pn \pt.v - 2p.\pt \pn.v  \big]
\end{align}
For $A=LR$,
\begin{align}
\frac{d^3\Gamma}{dx^3}\bigg|_{LR}^{(1)}=& ~ -\frac{4m_c }{m_b^2}\big( 2m_b^2+\lambda_1+3\lambda_2    \big) \pt.\pn \\[8pt]
\frac{d^3\Gamma}{dx^3}\bigg|_{LR}^{(2)}=& ~ \frac{8m_c}{m_b}\big[ -(\lambda_1+3\lambda_2)p.v\pt.\pn + m_b(\lambda_1+\lambda_2)\pt.\pn - 4m_b\lambda_2\pt.v\pn.v     \big] \\[8pt]
\frac{d^3\Gamma}{dx^3}\bigg|_{LR}^{(3)}=& ~ -\frac{32m_c \lambda_1 }{3} \big[ p.p - (p.v)^2   \big] \pt.\pn
\end{align}

For  $A=SLR$,
\begin{align}
\frac{d^3\Gamma}{dx^3}\bigg|_{SLR}^{(1)}=& ~\frac{m_{\tau}}{m_b^2}\big[ 2m_b^2(p.\pn + m_c\pn.v) + (\lambda_1+3\lambda_2)(p.\pn-m_b \pn.v)  \big] \\[8pt]
\frac{d^3\Gamma}{dx^3}\bigg|_{SLR}^{(2)}=& ~-\frac{2m_{\tau}}{3m_b}\big[ (\lambda_1+3\lambda_2)(-3p.\pn p.v+3m_b m_c\pn.v -5m_cp.v\pn.v + 2m_c p.\pn) \nonumber\\
										 & + m_b(5\lambda_1+3\lambda_2)p.\pn -2m_b(\lambda_1-3\lambda_2)p.v\pn.v   \big]\\[8pt]
\frac{d^3\Gamma}{dx^3}\bigg|_{SLR}^{(3)}=& ~\frac{8m_{\tau}\lambda_1}{3}\big[ p.p-(p.v)^2  \big] \big( p.\pn + m_c\pn.v   \big)
\end{align}

For $A=PLR$ we have,
\begin{align}
\frac{d^3\Gamma}{dx^3}\bigg|_{PLR}^{(i)}=& ~\frac{d^3\Gamma}{dx^3}\bigg|_{SLR}^{(i)}(m_c \to -m_c) \quad \quad i=1,2,3.
\end{align}

For $A=LT$,
\begin{align}
\frac{d^3\Gamma}{dx^3}\bigg|_{LT}^{(1)}=& ~ -48m_{\tau}m_c \big( \pn.v \big) \\[8pt]
\frac{d^3\Gamma}{dx^3}\bigg|_{LT}^{(2)}=& ~ -\frac{16m_{\tau}m_c}{m_b} \big[ (\lambda_1+3\lambda_2)(-2p.\pn + 5p.v\pn.v) - 3m_b(\lambda_1-\lambda_2)\pn.v     \big] \\[8pt]
\frac{d^3\Gamma}{dx^3}\bigg|_{LT}^{(3)}=& ~ -64m_{\tau}m_c\lambda_1  \big[ p.p - (p.v)^2  \big] (\pn.v)
\end{align}

For $A=RT$,
\begin{align}
\frac{d^3\Gamma}{dx^3}\bigg|_{RT}^{(1)}=& ~\frac{24m_{\tau}}{m_b^2} \big[ 2m_b^2 p.\pn + (\lambda_1+3\lambda_2)(p.\pn - m_b \pn.v)     \big] \\[8pt]
\frac{d^3\Gamma}{dx^3}\bigg|_{RT}^{(2)}=& ~ -\frac{16m_{\tau}}{m_b} \big[ -3(\lambda_1+3\lambda_2)p.\pn p.v + m_b(5\lambda_1-\lambda_2)p.\pn - 2m_b(\lambda_1+\lambda_2)p.v\pn.v     \big] \\[8pt]
\frac{d^3\Gamma}{dx^3}\bigg|_{RT}^{(3)}=& ~64m_{\tau}\lambda_1  \big[ p.p - (p.v)^2  \big]  \big( p.\pn \big) 
\end{align}

For $A=SPT$,
\begin{align}
\frac{d^3\Gamma}{dx^3}\bigg|_{SPT}^{(1)}=& ~8 \big(p.\pn \pt.v - p.\pt \pn.v   \big) \\[8pt]
\frac{d^3\Gamma}{dx^3}\bigg|_{SPT}^{(2)}=& ~\frac{8}{3m_b} \big[ 5(\lambda_1+3\lambda_2)p.v - m_b(5\lambda_1 + 3\lambda_2)  \big] \big(p.\pn \pt.v - p.\pt \pn.v   \big) \\[8pt]
\frac{d^3\Gamma}{dx^3}\bigg|_{SPT}^{(3)}=& ~\frac{32\lambda_1}{3} \big[ p.p -(p.v)^2  \big] \big(p.\pn \pt.v - p.\pt \pn.v   \big)
\end{align}

\newpage

\bibliography{inclusive}

\providecommand{\href}[2]{#2}\begingroup\raggedright\begin{thebibliography}{10}

\bibitem{Lees:2013uzd}
{\bfseries BaBar} Collaboration, J.~P. Lees {\em et al.}, ``{Measurement of an
  Excess of $\bar{B} \to D^{(*)}\tau^- \bar{\nu}_\tau$ Decays and Implications
  for Charged Higgs Bosons},''
  \href{http://dx.doi.org/10.1103/PhysRevD.88.072012}{{\em Phys. Rev.}
  {\bfseries D88} no.~7, (2013) 072012},
\href{http://arxiv.org/abs/1303.0571}{{\ttfamily arXiv:1303.0571 [hep-ex]}}.

\bibitem{Lees:2012xj}
{\bfseries BaBar} Collaboration, J.~P. Lees {\em et al.}, ``{Evidence for an
  excess of $\bar{B} \to D^{(*)}\tau^-\bar{\nu}_\tau $ decays},''
  \href{http://dx.doi.org/10.1103/PhysRevLett.109.101802}{{\em Phys. Rev.
  Lett.} {\bfseries 109} (2012) 101802},
\href{http://arxiv.org/abs/1205.5442}{{\ttfamily arXiv:1205.5442 [hep-ex]}}.

\bibitem{Huschle:2015rga}
{\bfseries Belle} Collaboration, M.~Huschle {\em et al.}, ``{Measurement of the
  branching ratio of $\bar{B} \to D^{(\ast)} \tau^- \bar{\nu}_\tau$ relative to
  $\bar{B} \to D^{(\ast)} \ell^- \bar{\nu}_\ell$ decays with hadronic tagging
  at Belle},'' \href{http://dx.doi.org/10.1103/PhysRevD.92.072014}{{\em Phys.
  Rev.} {\bfseries D92} no.~7, (2015) 072014},
\href{http://arxiv.org/abs/1507.03233}{{\ttfamily arXiv:1507.03233 [hep-ex]}}.

\bibitem{Abdesselam:2016cgx}
{\bfseries Belle} Collaboration, A.~Abdesselam {\em et al.}, ``{Measurement of
  the branching ratio of $\bar{B}^0 \rightarrow D^{*+} \tau^- \bar{\nu}_{\tau}$
  relative to $\bar{B}^0 \rightarrow D^{*+} \ell^- \bar{\nu}_{\ell}$ decays
  with a semileptonic tagging method},'' in {\em {Proceedings, 51st Rencontres
  de Moriond on Electroweak Interactions and Unified Theories: La Thuile,
  Italy, March 12-19, 2016}}.
\newblock 2016.
\newblock
\href{http://arxiv.org/abs/1603.06711}{{\ttfamily arXiv:1603.06711 [hep-ex]}}.
\newblock

\bibitem{Sato:2016svk}
{\bfseries Belle} Collaboration, Y.~Sato {\em et al.}, ``{Measurement of the
  branching ratio of $\bar{B}^0 \rightarrow D^{*+} \tau^- \bar{\nu}_{\tau}$
  relative to $\bar{B}^0 \rightarrow D^{*+} \ell^- \bar{\nu}_{\ell}$ decays
  with a semileptonic tagging method},''
  \href{http://dx.doi.org/10.1103/PhysRevD.94.072007}{{\em Phys. Rev.}
  {\bfseries D94} no.~7, (2016) 072007},
\href{http://arxiv.org/abs/1607.07923}{{\ttfamily arXiv:1607.07923 [hep-ex]}}.

\bibitem{Hirose:2016wfn}
{\bfseries Belle} Collaboration, S.~Hirose {\em et al.}, ``{Measurement of the
  $\tau$ lepton polarization and $R(D^*)$ in the decay $\bar{B} \to D^* \tau^-
  \bar{\nu}_\tau$},''
  \href{http://dx.doi.org/10.1103/PhysRevLett.118.211801}{{\em Phys. Rev.
  Lett.} {\bfseries 118} no.~21, (2017) 211801},
\href{http://arxiv.org/abs/1612.00529}{{\ttfamily arXiv:1612.00529 [hep-ex]}}.

\bibitem{Aaij:2015yra}
{\bfseries LHCb} Collaboration, R.~Aaij {\em et al.}, ``{Measurement of the
  ratio of branching fractions $\mathcal{B}(\bar{B}^0 \to
  D^{*+}\tau^{-}\bar{\nu}_{\tau})/\mathcal{B}(\bar{B}^0 \to
  D^{*+}\mu^{-}\bar{\nu}_{\mu})$},''
  \href{http://dx.doi.org/10.1103/PhysRevLett.115.159901,
  10.1103/PhysRevLett.115.111803}{{\em Phys. Rev. Lett.} {\bfseries 115}
  no.~11, (2015) 111803}, \href{http://arxiv.org/abs/1506.08614}{{\ttfamily
  arXiv:1506.08614 [hep-ex]}}.
[Erratum: Phys. Rev. Lett.115,no.15,159901(2015)].

\bibitem{Celis:2012dk}
A.~Celis, M.~Jung, X.-Q. Li, and A.~Pich, ``{Sensitivity to charged scalars in
  $\boldsymbol{B\to D^{(*)}\tau\nu_\tau}$ and $\boldsymbol{B\to\tau\nu_\tau}$
  decays},'' \href{http://dx.doi.org/10.1007/JHEP01(2013)054}{{\em JHEP}
  {\bfseries 01} (2013) 054},
\href{http://arxiv.org/abs/1210.8443}{{\ttfamily arXiv:1210.8443 [hep-ph]}}.

\bibitem{Duraisamy:2013kcw}
M.~Duraisamy and A.~Datta, ``{The Full $B \to D^{*} \tau^{-} \bar{\nu_\tau}$
  Angular Distribution and CP violating Triple Products},''
  \href{http://dx.doi.org/10.1007/JHEP09(2013)059}{{\em JHEP} {\bfseries 09}
  (2013) 059},
\href{http://arxiv.org/abs/1302.7031}{{\ttfamily arXiv:1302.7031 [hep-ph]}}.

\bibitem{Crivellin:2013wna}
A.~Crivellin, A.~Kokulu, and C.~Greub, ``{Flavor-phenomenology of
  two-Higgs-doublet models with generic Yukawa structure},''
  \href{http://dx.doi.org/10.1103/PhysRevD.87.094031}{{\em Phys. Rev.}
  {\bfseries D87} no.~9, (2013) 094031},
\href{http://arxiv.org/abs/1303.5877}{{\ttfamily arXiv:1303.5877 [hep-ph]}}.

\bibitem{Dorsner:2013tla}
I.~Doršner, S.~Fajfer, N.~Košnik, and I.~Nišandžić, ``{Minimally flavored
  colored scalar in $\bar B \to D^{(*)} \tau \bar \nu$ and the mass matrices
  constraints},'' \href{http://dx.doi.org/10.1007/JHEP11(2013)084}{{\em JHEP}
  {\bfseries 11} (2013) 084},
\href{http://arxiv.org/abs/1306.6493}{{\ttfamily arXiv:1306.6493 [hep-ph]}}.

\bibitem{Freytsis:2015qca}
M.~Freytsis, Z.~Ligeti, and J.~T. Ruderman, ``{Flavor models for $\bar{B} \to
  D^{(*)} \tau \bar{\nu}$},''
  \href{http://dx.doi.org/10.1103/PhysRevD.92.054018}{{\em Phys. Rev.}
  {\bfseries D92} no.~5, (2015) 054018},
\href{http://arxiv.org/abs/1506.08896}{{\ttfamily arXiv:1506.08896 [hep-ph]}}.

\bibitem{Deshpand:2016cpw}
N.~G. Deshpande and X.-G. He, ``{Consequences of R-parity violating
  interactions for anomalies in $\bar B\to D^{(*)} \tau \bar \nu$ and $b\to s
  \mu^+\mu^-$},'' \href{http://dx.doi.org/10.1140/epjc/s10052-017-4707-y}{{\em
  Eur. Phys. J.} {\bfseries C77} no.~2, (2017) 134},
\href{http://arxiv.org/abs/1608.04817}{{\ttfamily arXiv:1608.04817 [hep-ph]}}.

\bibitem{Bhattacharya:2018kig}
S.~Bhattacharya, S.~Nandi, and S.~Kumar~Patra, ``{$b \to c \tau \nu_{\tau}$
  Decays: A Catalogue to Compare, Constrain, and Correlate New Physics
  Effects},''
\href{http://arxiv.org/abs/1805.08222}{{\ttfamily arXiv:1805.08222 [hep-ph]}}.

\bibitem{Datta:2012qk}
A.~Datta, M.~Duraisamy, and D.~Ghosh, ``{Diagnosing New Physics in $b \to c \,
  \tau \, \nu_\tau$ decays in the light of the recent BaBar result},''
  \href{http://dx.doi.org/10.1103/PhysRevD.86.034027}{{\em Phys. Rev.}
  {\bfseries D86} (2012) 034027},
\href{http://arxiv.org/abs/1206.3760}{{\ttfamily arXiv:1206.3760 [hep-ph]}}.

\bibitem{Duraisamy:2014sna}
M.~Duraisamy, P.~Sharma, and A.~Datta, ``{Azimuthal $B \to D^{*} \tau^{-}
  \bar{\nu_\tau}$ angular distribution with tensor operators},''
  \href{http://dx.doi.org/10.1103/PhysRevD.90.074013}{{\em Phys. Rev.}
  {\bfseries D90} no.~7, (2014) 074013},
\href{http://arxiv.org/abs/1405.3719}{{\ttfamily arXiv:1405.3719 [hep-ph]}}.

\bibitem{Bhattacharya:2014wla}
B.~Bhattacharya, A.~Datta, D.~London, and S.~Shivashankara, ``{Simultaneous
  Explanation of the $R_K$ and $R(D^{(*)})$ Puzzles},''
  \href{http://dx.doi.org/10.1016/j.physletb.2015.02.011}{{\em Phys. Lett.}
  {\bfseries B742} (2015) 370--374},
\href{http://arxiv.org/abs/1412.7164}{{\ttfamily arXiv:1412.7164 [hep-ph]}}.

\bibitem{Bailey:2012jg}
J.~A. Bailey {\em et al.}, ``{Refining new-physics searches in $B \to D \tau
  \nu$ decay with lattice QCD},''
  \href{http://dx.doi.org/10.1103/PhysRevLett.109.071802}{{\em Phys. Rev.
  Lett.} {\bfseries 109} (2012) 071802},
\href{http://arxiv.org/abs/1206.4992}{{\ttfamily arXiv:1206.4992 [hep-ph]}}.

\bibitem{Lattice:2015rga}
{\bfseries MILC} Collaboration, J.~A. Bailey {\em et al.}, ``{ $B \to D \ell
  \nu$ form factors at nonzero recoil and |V$_{cb}$| from 2+1-flavor lattice
  QCD},'' \href{http://dx.doi.org/10.1103/PhysRevD.92.034506}{{\em Phys. Rev.}
  {\bfseries D92} no.~3, (2015) 034506},
\href{http://arxiv.org/abs/1503.07237}{{\ttfamily arXiv:1503.07237 [hep-lat]}}.

\bibitem{Na:2015kha}
{\bfseries HPQCD} Collaboration, H.~Na, C.~M. Bouchard, G.~P. Lepage,
  C.~Monahan, and J.~Shigemitsu, ``{$B \rightarrow D l \nu$ form factors at
  nonzero recoil and extraction of $|V_{cb}|$},''
  \href{http://dx.doi.org/10.1103/PhysRevD.93.119906,
  10.1103/PhysRevD.92.054510}{{\em Phys. Rev.} {\bfseries D92} no.~5, (2015)
  054510}, \href{http://arxiv.org/abs/1505.03925}{{\ttfamily arXiv:1505.03925
  [hep-lat]}}.
[Erratum: Phys. Rev.D93,no.11,119906(2016)].

\bibitem{Bigi:2016mdz}
D.~Bigi and P.~Gambino, ``{Revisiting $B\to D \ell \nu$},''
  \href{http://dx.doi.org/10.1103/PhysRevD.94.094008}{{\em Phys. Rev.}
  {\bfseries D94} no.~9, (2016) 094008},
\href{http://arxiv.org/abs/1606.08030}{{\ttfamily arXiv:1606.08030 [hep-ph]}}.

\bibitem{Bigi:2017jbd}
D.~Bigi, P.~Gambino, and S.~Schacht, ``{$R(D^*)$, $|V_{cb}|$, and the Heavy
  Quark Symmetry relations between form factors},''
  \href{http://dx.doi.org/10.1007/JHEP11(2017)061}{{\em JHEP} {\bfseries 11}
  (2017) 061},
\href{http://arxiv.org/abs/1707.09509}{{\ttfamily arXiv:1707.09509 [hep-ph]}}.

\bibitem{Bernlochner:2017jka}
F.~U. Bernlochner, Z.~Ligeti, M.~Papucci, and D.~J. Robinson, ``{Combined
  analysis of semileptonic $B$ decays to $D$ and $D^*$: $R(D^{(*)})$,
  $|V_{cb}|$, and new physics},''
  \href{http://dx.doi.org/10.1103/PhysRevD.95.115008,
  10.1103/PhysRevD.97.059902}{{\em Phys. Rev.} {\bfseries D95} no.~11, (2017)
  115008}, \href{http://arxiv.org/abs/1703.05330}{{\ttfamily arXiv:1703.05330
  [hep-ph]}}.
[Erratum: Phys. Rev.D97,no.5,059902(2018)].

\bibitem{Jaiswal:2017rve}
S.~Jaiswal, S.~Nandi, and S.~K. Patra, ``{Extraction of $|V_{cb}|$ from $B\to
  D^{(*)}\ell\nu_\ell$ and the Standard Model predictions of $R(D^{(*)})$},''
  \href{http://dx.doi.org/10.1007/JHEP12(2017)060}{{\em JHEP} {\bfseries 12}
  (2017) 060},
\href{http://arxiv.org/abs/1707.09977}{{\ttfamily arXiv:1707.09977 [hep-ph]}}.

\bibitem{HFLAV16}
{\bfseries Heavy Flavor Averaging Group} Collaboration, Y.~Amhis {\em et al.},
  ``{Averages of $b$-hadron, $c$-hadron, and $\tau$-lepton properties as of
  summer 2016},'' \href{http://dx.doi.org/10.1140/epjc/s10052-017-5058-4}{{\em
  Eur. Phys. J.} {\bfseries C77} (2017) 895},
  \href{http://arxiv.org/abs/1612.07233}{{\ttfamily arXiv:1612.07233
  [hep-ex]}}.
{updated results and plots available at
  \href{https://hflav.web.cern.ch}{{\texttt{https://hflav.web.cern.ch}}}}.

\bibitem{Kamali:2018fhr}
S.~Kamali, A.~Rashed, and A.~Datta, ``{New physics in inclusive $B \to X_c\ell
  \bar{\nu}$ decay in light of $R(D^{(*)})$ measurements},''
  \href{http://dx.doi.org/10.1103/PhysRevD.97.095034}{{\em Phys. Rev.}
  {\bfseries D97} no.~9, (2018) 095034},
\href{http://arxiv.org/abs/1801.08259}{{\ttfamily arXiv:1801.08259 [hep-ph]}}.

\bibitem{Grossman:1994ax}
Y.~Grossman and Z.~Ligeti, ``{The Inclusive $\bar{B} \to \tau \bar{\nu}_\tau X$
  decay in two Higgs doublet models},''
  \href{http://dx.doi.org/10.1016/0370-2693(94)91267-X}{{\em Phys. Lett.}
  {\bfseries B332} (1994) 373--380},
\href{http://arxiv.org/abs/hep-ph/9403376}{{\ttfamily arXiv:hep-ph/9403376
  [hep-ph]}}.

\bibitem{Colangelo:2016ymy}
P.~Colangelo and F.~De~Fazio, ``{Tension in the inclusive versus exclusive
  determinations of $|V_{cb}|$: a possible role of new physics},''
  \href{http://dx.doi.org/10.1103/PhysRevD.95.011701}{{\em Phys. Rev.}
  {\bfseries D95} no.~1, (2017) 011701},
\href{http://arxiv.org/abs/1611.07387}{{\ttfamily arXiv:1611.07387 [hep-ph]}}.

\bibitem{Manohar:1993qn}
A.~V. Manohar and M.~B. Wise, ``{Inclusive semileptonic B and polarized
  $\Lambda_b$ decays from QCD},''
  \href{http://dx.doi.org/10.1103/PhysRevD.49.1310}{{\em Phys. Rev.} {\bfseries
  D49} (1994) 1310--1329},
\href{http://arxiv.org/abs/hep-ph/9308246}{{\ttfamily arXiv:hep-ph/9308246
  [hep-ph]}}.

\bibitem{Balk:1993sz}
S.~Balk, J.~G. Korner, D.~Pirjol, and K.~Schilcher, ``{Inclusive semileptonic B
  decays in QCD including lepton mass effects},''
  \href{http://dx.doi.org/10.1007/BF01557233}{{\em Z. Phys.} {\bfseries C64}
  (1994) 37--44},
\href{http://arxiv.org/abs/hep-ph/9312220}{{\ttfamily arXiv:hep-ph/9312220
  [hep-ph]}}.

\bibitem{Falk:1993dh}
A.~F. Falk, M.~E. Luke, and M.~J. Savage, ``{Nonperturbative contributions to
  the inclusive rare decays $B \to X_s \gamma$ and $B \to X_s l^+ l^-$},''
  \href{http://dx.doi.org/10.1103/PhysRevD.49.3367}{{\em Phys. Rev.} {\bfseries
  D49} (1994) 3367--3378},
\href{http://arxiv.org/abs/hep-ph/9308288}{{\ttfamily arXiv:hep-ph/9308288
  [hep-ph]}}.

\bibitem{Koyrakh:1993pq}
L.~Koyrakh, ``{Nonperturbative corrections to the heavy lepton energy
  distribution in the inclusive decays $H(b) \to \tau \bar{\nu} X$},''
  \href{http://dx.doi.org/10.1103/PhysRevD.49.3379}{{\em Phys. Rev.} {\bfseries
  D49} (1994) 3379--3384},
\href{http://arxiv.org/abs/hep-ph/9311215}{{\ttfamily arXiv:hep-ph/9311215
  [hep-ph]}}.

\bibitem{Falk:1994gw}
A.~F. Falk, Z.~Ligeti, M.~Neubert, and Y.~Nir, ``{Heavy quark expansion for the
  inclusive decay $\bar{B} \to \tau \bar{\nu} X$},''
  \href{http://dx.doi.org/10.1016/0370-2693(94)91206-8}{{\em Phys. Lett.}
  {\bfseries B326} (1994) 145--153},
\href{http://arxiv.org/abs/hep-ph/9401226}{{\ttfamily arXiv:hep-ph/9401226
  [hep-ph]}}.

\bibitem{Blok:1993va}
B.~Blok, L.~Koyrakh, M.~A. Shifman, and A.~I. Vainshtein, ``{Differential
  distributions in semileptonic decays of the heavy flavors in QCD},''
  \href{http://dx.doi.org/10.1103/PhysRevD.50.3572,
  10.1103/PhysRevD.49.3356}{{\em Phys. Rev.} {\bfseries D49} (1994) 3356},
  \href{http://arxiv.org/abs/hep-ph/9307247}{{\ttfamily arXiv:hep-ph/9307247
  [hep-ph]}}.
[Erratum: Phys. Rev.D50,3572(1994)].

\bibitem{Ligeti:2014kia}
Z.~Ligeti and F.~J. Tackmann, ``{Precise predictions for $B \to X_c \tau \bar
  \nu$ decay distributions},''
  \href{http://dx.doi.org/10.1103/PhysRevD.90.034021}{{\em Phys. Rev.}
  {\bfseries D90} no.~3, (2014) 034021},
\href{http://arxiv.org/abs/1406.7013}{{\ttfamily arXiv:1406.7013 [hep-ph]}}.

\bibitem{Dassinger:2006md}
B.~M. Dassinger, T.~Mannel, and S.~Turczyk, ``{Inclusive semi-leptonic B decays
  to order $1/m_b^4$},''
  \href{http://dx.doi.org/10.1088/1126-6708/2007/03/087}{{\em JHEP} {\bfseries
  03} (2007) 087},
\href{http://arxiv.org/abs/hep-ph/0611168}{{\ttfamily arXiv:hep-ph/0611168
  [hep-ph]}}.

\bibitem{Hoang:1998ng}
A.~H. Hoang, Z.~Ligeti, and A.~V. Manohar, ``{B decay and the Upsilon mass},''
  \href{http://dx.doi.org/10.1103/PhysRevLett.82.277}{{\em Phys. Rev. Lett.}
  {\bfseries 82} (1999) 277--280},
\href{http://arxiv.org/abs/hep-ph/9809423}{{\ttfamily arXiv:hep-ph/9809423
  [hep-ph]}}.

\bibitem{Hoang:1998hm}
A.~H. Hoang, Z.~Ligeti, and A.~V. Manohar, ``{B decays in the upsilon
  expansion},'' \href{http://dx.doi.org/10.1103/PhysRevD.59.074017}{{\em Phys.
  Rev.} {\bfseries D59} (1999) 074017},
\href{http://arxiv.org/abs/hep-ph/9811239}{{\ttfamily arXiv:hep-ph/9811239
  [hep-ph]}}.

\bibitem{Benson:2003kp}
D.~Benson, I.~I. Bigi, T.~Mannel, and N.~Uraltsev, ``{Imprecated, yet
  impeccable: On the theoretical evaluation of Gamma(B ---> X(c) l nu)},''
  \href{http://dx.doi.org/10.1016/S0550-3213(03)00452-8}{{\em Nucl. Phys.}
  {\bfseries B665} (2003) 367--401},
\href{http://arxiv.org/abs/hep-ph/0302262}{{\ttfamily arXiv:hep-ph/0302262
  [hep-ph]}}.

\bibitem{Gambino:2004qm}
P.~Gambino and N.~Uraltsev, ``{Moments of semileptonic B decay distributions in
  the 1/m(b) expansion},''
  \href{http://dx.doi.org/10.1140/epjc/s2004-01671-2}{{\em Eur. Phys. J.}
  {\bfseries C34} (2004) 181--189},
\href{http://arxiv.org/abs/hep-ph/0401063}{{\ttfamily arXiv:hep-ph/0401063
  [hep-ph]}}.

\bibitem{Gambino:2011cq}
P.~Gambino, ``{B semileptonic moments at NNLO},''
  \href{http://dx.doi.org/10.1007/JHEP09(2011)055}{{\em JHEP} {\bfseries 09}
  (2011) 055},
\href{http://arxiv.org/abs/1107.3100}{{\ttfamily arXiv:1107.3100 [hep-ph]}}.

\bibitem{Alberti:2014yda}
A.~Alberti, P.~Gambino, K.~J. Healey, and S.~Nandi, ``{Precision Determination
  of the Cabibbo-Kobayashi-Maskawa Element $V_{cb}$},''
  \href{http://dx.doi.org/10.1103/PhysRevLett.114.061802}{{\em Phys. Rev.
  Lett.} {\bfseries 114} no.~6, (2015) 061802},
\href{http://arxiv.org/abs/1411.6560}{{\ttfamily arXiv:1411.6560 [hep-ph]}}.

\bibitem{Bauer:2002sh}
C.~W. Bauer, Z.~Ligeti, M.~Luke, and A.~V. Manohar, ``{B decay shape variables
  and the precision determination of |V(cb)| and m(b)},''
  \href{http://dx.doi.org/10.1103/PhysRevD.67.054012}{{\em Phys. Rev.}
  {\bfseries D67} (2003) 054012},
\href{http://arxiv.org/abs/hep-ph/0210027}{{\ttfamily arXiv:hep-ph/0210027
  [hep-ph]}}.

\bibitem{Bauer:2004ve}
C.~W. Bauer, Z.~Ligeti, M.~Luke, A.~V. Manohar, and M.~Trott, ``{Global
  analysis of inclusive B decays},''
  \href{http://dx.doi.org/10.1103/PhysRevD.70.094017}{{\em Phys. Rev.}
  {\bfseries D70} (2004) 094017},
\href{http://arxiv.org/abs/hep-ph/0408002}{{\ttfamily arXiv:hep-ph/0408002
  [hep-ph]}}.

\bibitem{Mannel:2017jfk}
T.~Mannel, A.~V. Rusov, and F.~Shahriaran, ``{Inclusive semitauonic $B$ decays
  to order ${\cal O}(\Lambda_{QCD}^3/m_b^3) $},''
  \href{http://dx.doi.org/10.1016/j.nuclphysb.2017.05.016}{{\em Nucl. Phys.}
  {\bfseries B921} (2017) 211--224},
\href{http://arxiv.org/abs/1702.01089}{{\ttfamily arXiv:1702.01089 [hep-ph]}}.

\bibitem{Aquila:2005hq}
V.~Aquila, P.~Gambino, G.~Ridolfi, and N.~Uraltsev, ``{Perturbative corrections
  to semileptonic b decay distributions},''
  \href{http://dx.doi.org/10.1016/j.nuclphysb.2005.04.031}{{\em Nucl. Phys.}
  {\bfseries B719} (2005) 77--102},
\href{http://arxiv.org/abs/hep-ph/0503083}{{\ttfamily arXiv:hep-ph/0503083
  [hep-ph]}}.

\bibitem{Jezabek:1996db}
M.~Jezabek and L.~Motyka, ``{Tau lepton distributions in semileptonic B
  decays},'' \href{http://dx.doi.org/10.1016/S0550-3213(97)00341-6}{{\em Nucl.
  Phys.} {\bfseries B501} (1997) 207--223},
\href{http://arxiv.org/abs/hep-ph/9701358}{{\ttfamily arXiv:hep-ph/9701358
  [hep-ph]}}.

\bibitem{Biswas:2009rb}
S.~Biswas and K.~Melnikov, ``{Second order QCD corrections to inclusive
  semileptonic $b \to X_c l \bar{\nu}_l$ decays with massless and massive
  lepton},'' \href{http://dx.doi.org/10.1007/JHEP02(2010)089}{{\em JHEP}
  {\bfseries 02} (2010) 089},
\href{http://arxiv.org/abs/0911.4142}{{\ttfamily arXiv:0911.4142 [hep-ph]}}.

\bibitem{Barate:2000rc}
{\bfseries ALEPH} Collaboration, R.~Barate {\em et al.}, ``{Measurements of
  $BR(b \to \tau^- \bar{\nu}_\tau X)$ and $BR (b \to \tau^- \bar{\nu}_\tau
  D^{*\pm} X)$ and upper limits on $BR (B^- \to \tau^- \bar{\nu}_\tau)$ and $BR
  (b \to s \nu \bar{\nu})$},''
  \href{http://dx.doi.org/10.1007/s100520100612}{{\em Eur. Phys. J.} {\bfseries
  C19} (2001) 213--227},
\href{http://arxiv.org/abs/hep-ex/0010022}{{\ttfamily arXiv:hep-ex/0010022
  [hep-ex]}}.

\bibitem{Aaij:2015bfa}
{\bfseries LHCb} Collaboration, R.~Aaij {\em et al.}, ``{Determination of the
  quark coupling strength $|V_{ub}|$ using baryonic decays},''
  \href{http://dx.doi.org/10.1038/nphys3415}{{\em Nature Phys.} {\bfseries 11}
  (2015) 743--747},
\href{http://arxiv.org/abs/1504.01568}{{\ttfamily arXiv:1504.01568 [hep-ex]}}.

\bibitem{Celis:2016azn}
A.~Celis, M.~Jung, X.-Q. Li, and A.~Pich, ``{Scalar contributions to $b\to c
  (u) \tau \nu$ transitions},''
  \href{http://dx.doi.org/10.1016/j.physletb.2017.05.037}{{\em Phys. Lett.}
  {\bfseries B771} (2017) 168--179},
\href{http://arxiv.org/abs/1612.07757}{{\ttfamily arXiv:1612.07757 [hep-ph]}}.

\bibitem{Li:2016vvp}
X.-Q. Li, Y.-D. Yang, and X.~Zhang, ``{Revisiting the one leptoquark solution
  to the R(D$^{(*)}$) anomalies and its phenomenological implications},''
  \href{http://dx.doi.org/10.1007/JHEP08(2016)054}{{\em JHEP} {\bfseries 08}
  (2016) 054},
\href{http://arxiv.org/abs/1605.09308}{{\ttfamily arXiv:1605.09308 [hep-ph]}}.

\bibitem{Alonso:2016oyd}
R.~Alonso, B.~Grinstein, and J.~Martin~Camalich, ``{Lifetime of $B_c^-$
  Constrains Explanations for Anomalies in $B\to D^{(*)}\tau\nu$},''
  \href{http://dx.doi.org/10.1103/PhysRevLett.118.081802}{{\em Phys. Rev.
  Lett.} {\bfseries 118} no.~8, (2017) 081802},
\href{http://arxiv.org/abs/1611.06676}{{\ttfamily arXiv:1611.06676 [hep-ph]}}.

\bibitem{Datta:2017aue}
A.~Datta, S.~Kamali, S.~Meinel, and A.~Rashed, ``{Phenomenology of $
  {\Lambda}_b\to {\Lambda}_c\tau {\overline{\nu}}_{\tau } $ using lattice QCD
  calculations},'' \href{http://dx.doi.org/10.1007/JHEP08(2017)131}{{\em JHEP}
  {\bfseries 08} (2017) 131},
\href{http://arxiv.org/abs/1702.02243}{{\ttfamily arXiv:1702.02243 [hep-ph]}}.

\end{thebibliography}\endgroup

\end{document}